\documentclass[aps,prb,twocolumn,citeautoscript]{revtex4-1}
\usepackage{amsmath,amssymb,mathrsfs,bm,feynmf,setspace}
\usepackage{natbib}
\usepackage{graphicx}
\usepackage[tight]{subfigure}
\usepackage{color}
\usepackage{hyperref}
\newcommand{\al}{\alpha}
\newcommand{\be}{\beta}

\newcommand{\de}{\delta}

\newcommand{\s}{\sigma}

\newcommand{\G}{\Gamma}

\newcommand{\beq}{\begin{equation}}
\newcommand{\eeq}{\end{equation}}
\newcommand{\Beq}{\begin{eqnarray}}
\newcommand{\Eeq}{\end{eqnarray}}
\newcommand{\bml}{\begin{multline}}

\newcommand{\eeqm}{\end{multline}}

\newcommand{\bsp}{\begin{split}}
\newcommand{\esp}{\end{split}}
\newcommand{\down}{\downarrow}
\newcommand{\up}{\uparrow}

\renewcommand{\b}[1]{{\bm #1}}

\newcommand{\req}[1]{Eq.~(\ref{#1})}

\DeclareMathOperator{\sgn}{sgn}

\newcommand{\toxy}{t_{\rm oxy}}

\begin{document}

\title{Topological and magnetic phases of interacting electrons in the pyrochlore iridates}
\author{William Witczak-Krempa$^1$ and Yong Baek Kim$^{1,2}$}
\affiliation{
$^1$Department of Physics, The University of Toronto, Toronto, Ontario M5S 1A7, Canada\\
$^3$School of Physics, Korea Institute for Advanced Study, Seoul 130-722, Korea 
} 
\date{\today}

\begin{abstract}
We construct a model for interacting electrons with strong spin orbit
coupling in the pyrochlore iridates. We establish the importance of
the direct hopping process between the Ir atoms and use the relative
strength of the direct and indirect hopping as a generic tuning
parameter to study the correlation effects across the iridates family.
We predict novel quantum phase transitions between conventional and/or
topologically non-trivial phases.
At weak coupling, we find topological insulator and metallic phases.
As one increases the interaction strength, various magnetic orders
emerge. The novel topological Weyl semi-metal phase is found to be realized
in these different orders, one of them being the all-in/all-out pattern. Our findings establish the possible magnetic ground states for
the iridates and suggest the generic presence of the Weyl semi-metal
phase in correlated magnetic insulators on the pyrochlore lattice.
We discus the implications for existing and future experiments.
\end{abstract}
\maketitle
\section{Introduction}
Topological insulators\cite{RMP_TI,ti_rev_zhang,moore-hasan} (TIs) have provided theorists and experimentalists alike with a new family of 
topologically non-trivial systems. In these materials, a sufficiently strong spin orbit coupling (SOC) leads to a peculiar band structure 
that cannot be adiabatically deformed to that of a flat band insulator without closing the bulk gap. This leads to robust boundary states 
that display momentum-spin locking. 
The materials in which these gapless helical surface states have been observed are weakly interacting semiconductors, for which the 
above theory was constructed.
An inviting question, therefore, relates to the kinds of quantum ground states that would arise 
in the presence of interactions in these systems or to interaction-driven TIs. 
For instance, several studies examined various kinds of fractionalized TIs\cite{pesin,will,levin,lehur,maciejko,swingle,qi}. 

In this context, transition metal oxides with 5$d$ transition metal elements 
may be ideal systems to search for TIs and new topological
phases in the presence of interactions. In these systems, the strength of the 
interaction and that of the SOC are comparable, providing a
playground for the interplay between two effects. 
In particular, the pyrochlore iridates, $A_2$Ir$_2$O$_7$, have been suggested to host various
topologically non-trivial states\cite{pesin,wan,balents-kiss,will,fiete-rev,fiete-trig}.
Here, $A$ is a Lanthanide or Yittrium, whose size affects the effective bandwidth of the 
5$d$ electrons of Ir via the Ir-O-Ir bond angle, thereby tuning the effective strength of the interaction.
Experiments on these compounds reveal metal-insulator transitions upon variation of temperature or chemical\cite{maeno} and external\cite{fazel} pressure, as well as indications of magnetism\cite{taira,musr}. 
%Although a direct probe of magnetism still awaits, a recent $\mu$SR measurement\cite{musr} on Eu$_2$Ir$_2$O$_7$ suggests the presence of long-range commensurate order.

In this work, we present a Hubbard-type model for the interacting electrons
in the pyrochlore iridates and determine the ground state phase diagram using mean 
field and strong coupling methods. We find that it is important to include both the indirect hopping
of $5d$ electrons of Ir through oxygens and the direct hopping between Ir sites.
This is because the $5d$ orbitals of Ir are spatially extended and
the nature of the ground state is sensitive to the relative strength of these
hopping amplitudes.
In the weakly interacting limit, both TIs and (semi-)metallic states are realized depending on
the relative strength of different hopping amplitudes. This is in contrast to
a previous work\cite{pesin} where only the indirect hopping process was considered
and only the TI phase was obtained in the large SOC limit (with the ideal
cubic crystal field). 
%The transition between the
%TI and (semi-)metal occurs ressembles to the one driven
%by trigonal crystal fields\cite{bj-trig}.
The interactions between electrons lead to two different magnetically ordered
ground states in different parameter regions. 
In particular, for intermediate interactions, the 
topological semi-metal\cite{abrikosov,*nielsen,*volovik-book,wan,balents-kiss} (TSM) state with Weyl-like fermions appears in both kinds of
AF phases. 
%A previous LSDA study\cite{wan} of
%the pyrochlore iridates obtained the TSM state with the
%all-in/all-out order.  
Our results suggest that the TSM state
and the related Mott insulating state can have different magnetic ordering 
patterns depending on the choice of the $A$-site ion or upon application of hydrostatic pressure,
leading to the possibility of novel quantum phase transitions in the iridates.

\section{Model and approach} 
In the atomic limit, the oxygen octahedra surrounding the Ir$^{4+}$ ions
create large cubic crystal fields that split the $5d$ orbitals into $t_{2g}$ and $e_g$ multiplets.
The five $5d$ electrons of Ir$^{4+}$ occupy the $t_{2g}$ levels, leaving the high energy 
$e_g$ levels empty. The angular momentum operator projected into the $t_{2g}$ levels
is effectively $\ell = 1$ with an extra negative sign, i.e.,
$P_{t_{2g}} \b L P_{t_{2g}} = - \b L^{\rm eff}_{\ell =1}$. 
The on-site SOC leads to a further splitting
into an effective pseudospin $j_{\rm eff} =1/2$ doublet and a $j_{\rm eff} = 3/2$ quadruplet,
the former lying higher in energy\cite{SrIrO-prl,*SrIrO-science}. 
For sufficiently large SOC, the half-filled $j_{\rm eff}=1/2$ doublets form a low energy
manifold as the fully occupied $j_{\rm eff}=3/2$ levels are sufficiently far from 
the Fermi level. 
\begin{figure}
\centering
\subfigure[]{\label{fig:pyro} \includegraphics[scale=0.3]{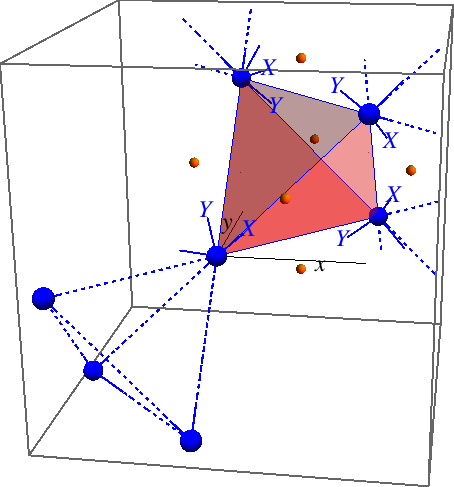}}
\subfigure[]{\label{fig:hop-pd} \includegraphics[scale=0.26]{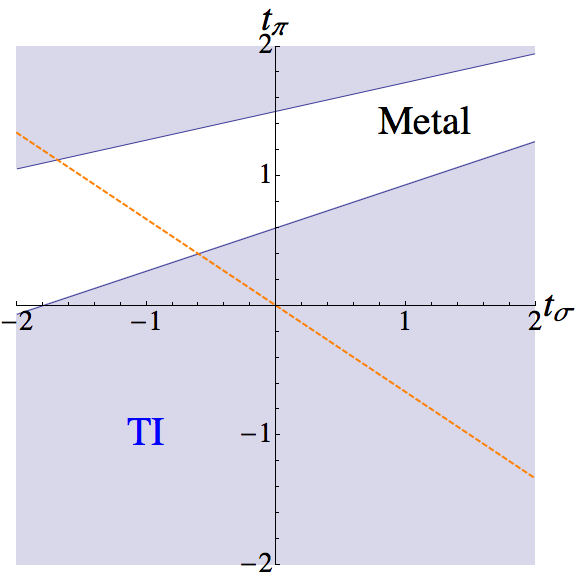}}
\caption{\label{fig:crystal_hop-pd} a) Pyrochlore lattice of Ir atoms (large).
 The oxygens (small) bridging the Ir's in one unit cell are shown together with the local axes they define. b) Phase diagram of the hopping Hamiltonian $H_0$. We set $\toxy=1$. The dashed line is $t_\pi=-2t_\s/3$. }
\end{figure}
In going to a tight-binding description, we need to take into account the different orientations of the local oxygen
octahedra at each of the 4 sites in the unit cell (see Fig.~\ref{fig:pyro}).
Previous studies\cite{pesin,bj-trig,fiete-trig} considered nearest neighbour Ir-Ir hopping 
mediated by the oxygens. In this work, we also include the direct hopping between
the Ir atoms, which is expected to be significant due to the large spatial extent 
of Iridium's $5d$ orbitals. 
%The direct hopping amplitudes are treated within the 
%Slater-Koster formalism. 
We 
consider only the $\pi$- and $\s$-overlaps between 
the $t_{2g}$ orbitals, neglecting the usually smaller 
$\delta$-overlap. This leaves us with \emph{two} direct hopping 
parameters: $t_\sigma$ and $t_\pi$. The resulting kinetic Hamiltonian reads 
\begin{equation}
%  \sum_{Ri\al}(\epsilon_\al-\mu)d_{Ri\al}^\dag d_{Ri\al}\\
 % +\sum_{\substack{
    %  \langle Ri, R'i'\rangle \\
 H _0=     \sum_{\substack{
    \langle Ri, R'i'\rangle, \al\al' }} (T_{\rm o,\al\al'}^{ii'}+T_{\rm
    d,\al\al'}^{ii'})d_{Ri\al}^\dag d_{R'i'\al'} \, ,
\end{equation}
where $R$ denotes the sites of the underlying Bravais 
FCC lattice of the pryrochlore lattice of Ir's, while $i=1,\dots,4$ labels
the sites within the unit cell. The operator $d_{Ri\up(\down)}$ annihilates an electron in the 
pseudospin $\up(\down)$ state at 
site $Ri$. The two sets of matrices $T_{\rm o}$ and $T_{\rm d}$ correspond to
the oxygen mediated\cite{pesin} and direct hopping, respectively. 

We include interactions via an on-site Hubbard repulsion between Iridium's $d$-electrons:
\begin{align}\label{eq:H}
  H &= H_0 + H_U , \\
  H_U &=U\sum_{Ri}n_{Ri\up}n_{Ri\down},
\end{align}
where $n_{Ri\al}$ is the density of
electrons occupying the $|j_{\rm eff} =1/2,\al\rangle$ state at site $Ri$, with
$\al=\up,\down$. As we are interested in the magnetic phases expected
at finite $U$, we perform a Hartree-Fock mean-field decoupling 
$H_U\rightarrow -U\sum_{Ri} (2\langle \bm j_{Ri}\rangle\cdot   
\bm j_{Ri} - \langle \bm j_{Ri}\rangle^2)$,
where $\bm j_{Ri}=\sum_{\al\be=\up,\down}d_{Ri\al}^\dag \bm
\sigma_{\al\be} d_{Ri\be}/2$ is the pseudospin operator, whose
expectation value will be determined self-consistently. We consider
magnetic configurations preserving the unit cell so that $\langle \bm j_{Ri}\rangle=
\langle \bm j_i\rangle$, $i=1,\dots, 4$, are the 4 order parameters under consideration.
These are directly proportional to the local magnetic moment carried by the
$d$-electrons. This follows from the fact that the projections of the spin and
orbital angular momentum operators onto the $j_{\rm eff}=1/2$ manifold 
are proportional to the pseudospin operator:
$\tilde P^\dag\b S \tilde P =-\b j/3$ and $\tilde P^\dag\b L \tilde P =-4\bm j/3$ with
$\tilde P = P_{t_{2g}} P_{1/2}$, where $P_{t_{2g}}$ projects onto the
$t_{2g}$ subspace and $P_{1/2}$ projects onto the $j_{\rm eff}=1/2$
subspace. This allows us to treat $\langle \bm j_i\rangle$ as the spontaneous local
magnetic moment of the electrons. 
%Indeed, we project the total magnetic moment of the electrons, $2\bm S+\bm L$, 
%on the $j=1/2$ manifold using the projection operator
%$\tilde P = P_{1/2}P_{t_{2g}}$, where $P_{t_{2g}}$ projects onto the
%$t_{2g}$ manifold and $P_{1/2}$ projects onto the $j=1/2$
%manifold. This leads simply to 
%\eq{
%  \tilde P (2\bm S+\bm L)\tilde P\inv = -2\bm j,
%} 
%as can be seen from the individual projections of the spin and
%angular momentum operators: $\tilde P\bm S \tilde P\inv =-\bm j/3$ and
% $\tilde P\bm L \tilde P\inv =4\bm j/3$. We shall not worry about the constant
% of proportionality and treat $\langle \bm j_i\rangle$ as the spontaneous net magnetic
% moment of the electrons. 
\section{Phase diagram}
\subsection{Metal and topological insulator at $U=0$}

We first examine the model at $U=0$. Fig.~\ref{fig:hop-pd} shows the resulting
phase diagram in terms of $t_\sigma$ and $t_\pi$ (we set $\toxy=1$ throughout). 
Notice that both insulating and metallic phases exist.
By virtue of the inversion
symmetry of the crystal, we use the Fu-Kane
formulas\cite{z2_fu-kane} for the $Z_2$ invariants in terms of the parity eigenvalues of the occupied
states at the time reversal invariant momenta (TRIMs) to determine the topological class of each insulating phase. 
We find that both insulating phases are TIs with indices $(1;000)$.
The TI phase adiabatically connected to $t_\sigma=t_\pi=0$ corresponds to the large spin orbit limit of 
Ref.\onlinecite{pesin} and is robust to the inclusion of weak direct hopping. As one tunes
the direct hoppings, a metallic phase eventually appears by means of a gap closing 
at the $\Gamma$ point. In the metal, the degeneracies at $\G$ become 2-4-2
compared to 4-2-2 in the TI (with time-reversal and inversion symmetries all band are doubly degenerate).  
A similar situation occurs in Refs.\onlinecite{bj-trig,fiete-trig}, where a trigonal distortion of the oxygen 
octahedra drives the transition,
not direct hopping as is the case here. The metallic phase is strictly speaking a semi-metal characterized by a point Fermi surface.  
Finite pockets can be generated by including very weak NNN hopping, as we
have explicitly verified. 
%However, this need not be the case: depending on the size and sign of the NNN hopping, the 
%semi-metal can remain. 
Although we don't consider trigonal distortions here, the direct hoppings alone can 
lead to qualitatively similar effects, {\it e.g.} the metallic phase resulting from the change 
in degeneracies at the $\G$ point.  
\subsection{Magnetic and topological phases at $U>0$}

We now turn to the $U>0$ case. For convenience, we restrict our attention to a one-dimensional 
cut in the $(t_\sigma,t_\pi)$ space defined by $t_\pi=-2t_\sigma/3$, as shown in Fig.~\ref{fig:hop-pd}. This is physically motivated since 
we expect $t_\sigma$ and $t_\pi$ to have opposite signs, with the $\s$-overlap being the strongest. Moreover,
the cut is representative as it intersects all the phases. In obtaining the finite $U$ diagram, we performed an
unconstrained analysis sampling over the space of all possible magnetic configurations preserving the unit cell.
% In obtaining the finite $U$ diagram, we considered different magnetic orders and compared their energies. In our analysis, we kept the magnitude of the order parameters at the 4 basis sites independent, resulting in 4 Ising order parameters. 
%We examined non-collinear orders, as well as AF and ferromagnetic (FM) collinear states along different
%directions. The non-collinear states considered are: 1) all-in/all-out, corresponding to all moments pointing towards (away from)
%the center of each tetrahedron; 2) 2-in/2-out, i.e. the spin-ice state; 3) 3-in/1-out. For the last two, we considered different
%configurations corresponding to various subsets of the 4 pseudospins pointing in or out. In the AF and FM, all pseudospins are collinearly aligned
%along (100), (010), (001), (111) or (110). The collinear AF states we considered all have zero net moment: 
%two pseudospins point along the chosen direction, while the other two are antiparallel.  
%It is important to note that all of the above orders preserve the unit cell, hence inversion.
\begin{figure}
\centering
\includegraphics[scale=0.3]{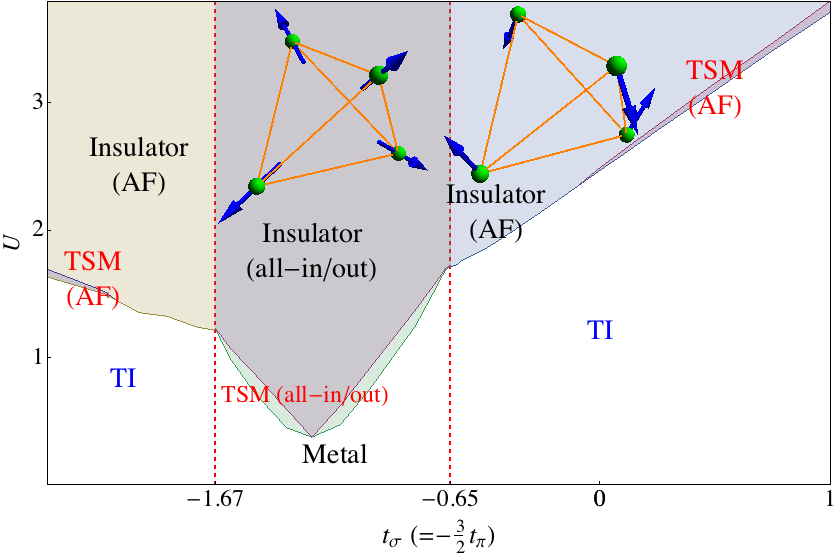}
\caption{\label{fig:Upd} Mean-field phase diagram ($\toxy=1$) as a function of $U$, the Hubbard coupling, and
the direct hopping parameters. The magnetic transitions from the TIs (metal) are 1st (2nd) order. }
\end{figure}
\begin{figure}%
\centering
   \includegraphics[scale=0.33]{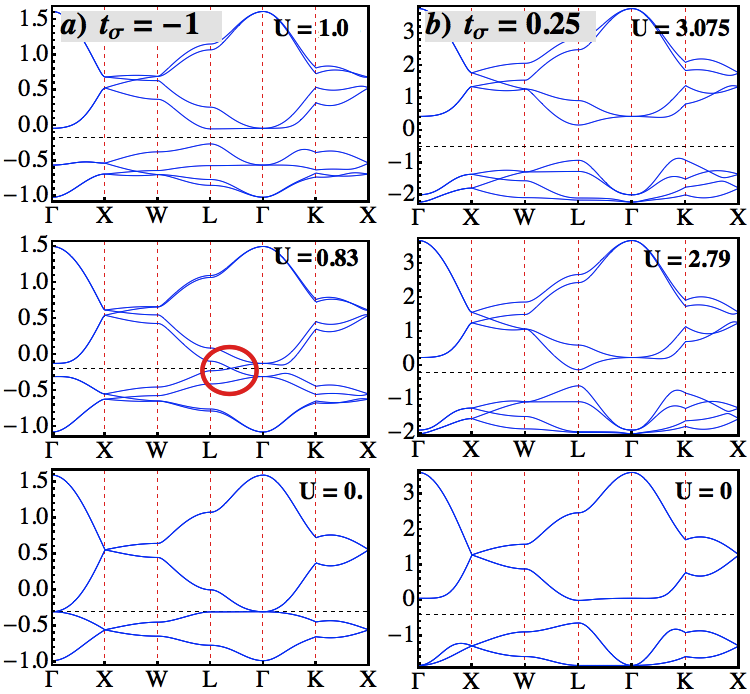}
  \caption{Evolution of the spectrum as a function of $U$. At intermediate $U$, in a), we can see a Weyl point along the
  $\G-L$ line, while in b), the spectrum naively seems insulating because the Weyl points lie away from high symmetry $k$-points. 
  The dashed line is the Fermi level.}
  \label{fig:edge}
\end{figure}

The resulting ground-state phase diagram appears in Fig.~\ref{fig:Upd}.
%The salient features of the resulting phase diagram (see Fig.~\ref{fig:Upd}) are: the nature of the transition, the resulting magnetic order and quasiparticle spectrum vary whether the non-interacting parent phase is a TI or metal. 
First, we note that the TI is more resilient to the magnetic instability than the metal, as expected due to the presence of the bulk gap in the former. Second, the magnetic phase transition resulting from increasing $U$ in the metal (TI) is second (first) order. Also, the magnetic order emerging from the TIs differs from the one found upon increasing $U$ in the metal. In the latter case, we find an all-in/out configuration while in the former the ground state is 3-fold degenerate (modulo the trivial degeneracy $\b j \rightarrow -\b j$): all 3 states result from the all-in/out state by performing $\pi/2$-rotations on the moments in the unit cell. These rotations occur within either one of the planes bisecting the 3 triangles meeting at each corner of the tetrahedron. The order emergent in both TI states is the same. In section \ref{sec:strong_coupling}, we discuss how the different magnetic orders and
the position of the transitions are actually connected to the corresponding ordering in the spin model obtained at large $U$: as $t_\sigma$ is tuned,
the induced Dzyaloshinski-Morya interaction alternates between the only two symmetry allowed possibilities on the pyrochlore lattice, leading to different
ordering.

\subsection{Topological Semi-metal}
By examining the spectra of the ordered phases, we discover that the so-called
topological semi-metal (TSM) is realized\footnote{We have included very weak NNN hopping to obtain the TSM in the all-in/out phase, for without it the cones are tilted such that there are lines at the Fermi level.} in the range $t_\s \geq -1.67$ and for a finite window of $U$. This semi-metallic phase has a Fermi ``surface" composed of points, each with a linearly dispersive spectrum of Weyl or two-component fermions, and may be considered as a 3D version of the Dirac points of graphene. The Hamiltonian near one such Weyl point takes the form
\beq
	H = \b v_0\cdot \b q + \sum_{i=1}^3 \b v_i\cdot \b q\sigma_i \, ,
\eeq
where $\b q=\b k-\b k_0$ is the deviation from the Weyl point at $\b k_0$. The Pauli matrices $\sigma_i$ represent the two bands involved in the touching, not (pseudo)spin.
One can assign a chiral ``charge" to these fermions, via the triple product of the 3 velocities: $c=\sgn(\b v_1\cdot \b v_2\times \b v_3)$. The massless nature of the two-component Weyl fermions is robust against local perturbations, which is not the case in 2D. As explained in Ref.\onlinecite{wan}, the only way to introduce a gap 
is to make two Weyl fermions with opposite chirality meet at some point in the BZ.
For this reason they are topological objects (see also the discussion below regarding the surface states). Further details relating to
the TSM can be found in Refs.~\onlinecite{abrikosov,nielsen,volovik-book,wan,balents-kiss,ran,burkov}.
 
The TSM appears in for both AF orders. In both cases we find a total of 8 Weyl points coming necessarily in 4 inversion-symmetry related pairs. The location and migration of these Weyl points depends on the magnetic order. Let us first examine the TSM phase present in the all-in/out state. In this case, the 8 Weyl points are born out of the quadratic touching at the $\G$ point as
the local moments spontaneously and continuously acquire a finite value with increasing $U>U_c$. Each pair of Weyl points lies on one of the four high symmetry lines joining $\Gamma$ to the four $L$ points, as can be seen in Fig.~\ref{fig:edge}. For this reason we only get 8 touchings, in contrast
to Ref.\onlinecite{wan}, where 24 Weyl points are obtained. In their case they live off the high symmetry lines so that each point is tripled by the
3-fold rotational symmetries about the $\Gamma-L$ lines. 
%We use the FCC reciprocal lattice vectors $\b G_1=\pi(-1,1,1),\b G_2=\pi(1,-1,1),\b G_3=\pi(1,1,-1)$, so that the 4 $L$ points are given by $\frac{1}{2}\b G_{i=1,2,3}$ and $\frac{1}{2}(\b G_1+\b G_2+\b G_3)$. The locations of the 8 Weyl points correspond to $\pm \ka\b G_{i=1,2,3}$ and $\pm \ka(\b G_1+\b G_2+\b G_3)$, where a single number $\ka$ controls
%the positions of all of the Weyl points. Note that $\ka=\ka(\Delta)$ increases monotonously from zero as the Ising magnetic order parameter, $\Delta=|\langle \b j_1\rangle|=\dots=|\langle \b j_4\rangle|$,
%goes from zero to a finite value.  
Weyl points of opposite chirality annihilate at the 4 $L$ points as $U$ is increased.
%There are two pairs that remain on the $\ka_3=0$
%plane, another pair lies along the $\b\ka=(0,0,1)$ direction and the final pair is bound to the $\b\ka=(1,1,1)$ line. 
%Explicitly, the location of the Dirac points are
%\begin{align}
%	1)& \;(\ka_*,0,0)\quad \&\quad (-\ka_*,0,0) \\
%	2)& \;(0,\ka_*,0)\quad \&\quad (0,-\ka_*,0) \\
%	3)& \;(0,0,\ka_*)\quad \&\quad (0,0,-\ka_*) \\
%	4)& \;(\ka_*,\ka_*,\ka_*)\quad \&\quad -(\ka_*,\ka_*,\ka_*)
%\end{align}
% where $\ka_*=\ka_*(\Delta)$ vanishes when the magnetic order parameter vanishes $\Delta=0$. The same $\ka_*$ determines the location
% of all the Dirac points. The Dirac points of opposite chirality are seen to annihilate each other when $\ka_*=-\ka_*$, i.e. when $\ka_*=1/2$. This corresponds to the four $L$ points:  
% \begin{align}
%	1)&\; \frac{1}{2}\b G_1=\pi(-\hat x+\hat y+\hat z) \\
%	2)&\;  \frac{1}{2}\b G_2=\pi(\hat x-\hat y+\hat z) \\
%	3)&\;  \frac{1}{2}\b G_3=\pi(\hat x+\hat y-\hat z) \\
%	4)&\; \frac{1}{2}(\b G_1+\b G_2+\b G_3) =\pi(\hat x+\hat y+\hat z)
%\end{align}
As they annihilate and create a gap, the parities of the highest
occupied states at these TRIMs change sign. 

Let us now consider the TSM arising from the TI, where we again have 8 Weyl points. The major difference is that they do not 
occur along high symmetry lines, as can be seen in Fig.~\ref{fig:edge}. We do not get 24 Weyl points because the magnetic order breaks the 3-fold rotational symmetries, which are preserved by the all-in/out state. We have explicitly located the Weyl points by looking at both the spectrum and density of states,
which shows a characteristic $(E-E_F)^2$ scaling. 
 %The resulting positions are $\pm \b k_a$ and $\pm \b k_a'$ for $a=1,2$, where
 %$\b k_a=\ka_1^{(a)}(\b G_1+\b G_3)+\ka_2 \b G_2$ and $\b k_a'=-\ka_1^{(a)}\b G_1+\ka_1^{(\bar a)} \b G_3$, with $\bar 1=2,\bar 2=1$.
% Upon closer inspection, we see that they can be divided into two categories according to whether they lie
% in the plane a) $\ka_1=\ka_3$ or b) $\ka_2=0$. More specifically: 
%\begin{align}
%	a.1)& \; (\ka_{*3}^{(1)},\ka_{*2},\ka_{*3}^{(1)}) \\
%	a.2)& \; (\ka_{*3}^{(2)},\ka_{*2},\ka_{*3}^{(2)}) \\
%	b.1)& \; (-\ka_{*3}^{(2)}, 0,\ka_{*3}^{(1)}) \\
%	b.2)& \; (-\ka_{*3}^{(1)}, 0,\ka_{*3}^{(2)}) 
%\end{align}
%The locations of the 8 Weyl points are governed by 3 numbers:
%$ \ka_1^{(1,2)}$ and $ \ka_2$. They annihilate in pairs $\pm \b k_1 \leftrightarrow \pm \b k_2$ and $\pm \b k_1' \leftrightarrow \pm \b k_2'$. For instance, this happens when  $\ka_1^{(1)}\nearrow 0.40$ and $\ka_1^{(2)}\searrow 0.40$ for $t_\s=0$.
The Weyl points don't annihilate at TRIMs, in contrast to the non-collinear TSM. As a result there is no parity flip associated with
the termination of the TSM phase when, upon increasing $U$, the system becomes insulating. 
\begin{figure}
\centering
\includegraphics[scale=0.33]{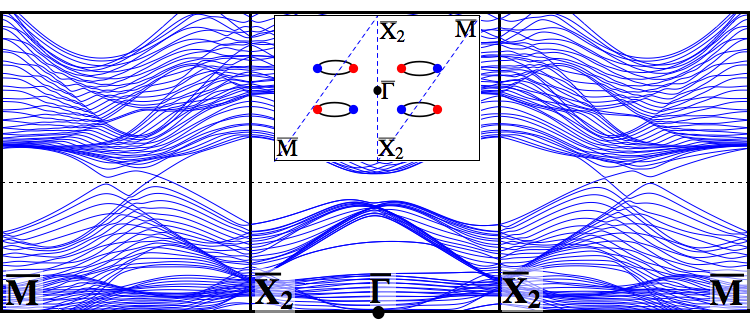}
\caption{\label{fig:ss} (Colors online) Spectrum for a slab that is finite along the (110) direction, with $t_\sigma=0.25$ and 
$U=2.8$ which corresponds to a TSM. The inset shows
the 2D BZ including the Fermi arcs connecting the projections of the Weyl points, where half of the Fermi arcs are located on the top surface, while the rest on the bottom one. The blue/red (dark/light) points correspond to Weyl bulk points with chiral charge $+1/-1$. }
\end{figure}

\emph{Surface states}:
The non-trivial band topology of the TSM (each Weyl point is a monopole of the U(1) Berry connection) leads to chiral surface states on certain surfaces, in analogy with the TI. In contrast with the latter,
the surface states of the former do not form closed Fermi surfaces, but rather open Fermi arcs. As argued in Ref.\onlinecite{wan}, the Fermi arcs join
the projections of bulk Weyl points of opposite chirality. As bulk Weyl points forming a pair are made to move towards each other by increasing $U$,
the corresponding Fermi arc shrinks, collapses to a point and disappears. 

In the TSM found at large $t_\s$, which we use to illustrate the Fermi arcs, there are no surface states along surfaces perpendicular to the (100), (010) or (001) directions.
For these surfaces, the projection process onto the 2D BZ maps 3D Weyl points of opposite chirality onto
the same 2D $k$-point. This leads to the absence of gapless surface states emanating
from the 2D $k$-point in question. 
%They related the number of surface states intersecting a loop in the 2D BZ to the Chern
%number of the 2D band structure defined on the surface of the cylinder generated by translating the loop in the 3D BZ along the normal to the surface. The corresponding Chern number is then equal to the enclosed net monopole charge of the $U(1)$ Berry connection, i.e. the sum of the chiralities of the enclosed Weyl points. 
For a surface perpendicular to the (110) direction, however, 
the projection is injective
and Fermi arcs exist, as we illustrate in Fig.~\ref{fig:ss}.

\section{Strong coupling expansion}
\label{sec:strong_coupling}

\begin{figure}
\centering
\subfigure[]{\label{fig:DM_vectors} \includegraphics[scale=0.31]{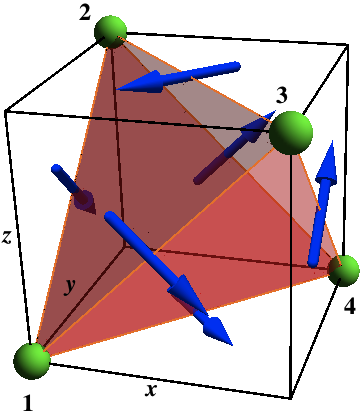}}
\subfigure[]{\label{fig:DM_sign} \includegraphics[scale=0.34]{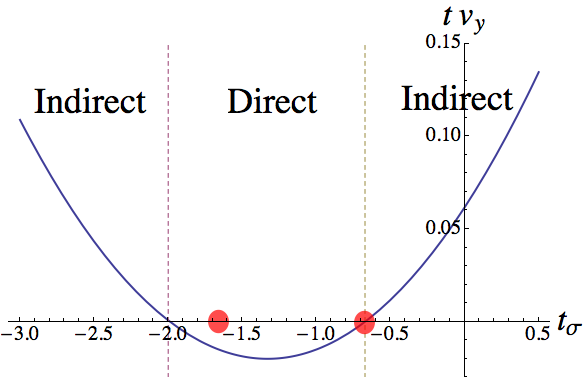}}
\caption{\label{fig:DM} a) $\b D$ vectors corresponding to the ``indirect" DM interaction. The ``direct"
type is obtained by changing the sign of all vectors. b) 
$tv_y$ as a function of the hopping strength $t_\sigma$. The sign of $tv_y$ determines the 
nature of the DM interaction: $tv_y>0$ $(tv_y<0)$ corresponds to the ``indirect" (``direct") type.
The red dots correspond to the metal-insulator transitions
at $U=0$, which coincide with the points at which the nature of the magnetic order changes for $U$ above the
ordering threshold.  }
\end{figure}

In this section, we discuss the large $U$ limit of our Hubbard Hamiltonian \req{eq:H}.
We show how the effective spin-1/2 model obtained in that limit sheds light
on the orders found in the mean field calculation as well as on the 
location of the phase transitions. In taking the limit where $U$ is much larger than all hopping amplitudes 
($t_{\rm oxy},t_\sigma,t_\pi$), we can use second order perturbation theory
to obtain the low energy spin Hamiltonian:
\begin{align}\label{eq:spin-hamil}
  H'=\sum_{ij}\left[J\b S_i\cdot\b S_j+\b D_{ij}\cdot(\b S_i\times \b S_j)
  + S_i^a \Gamma_{ij}^{ab} S_j^b \right] 
\end{align}
where the terms are, in order: the AF Heisenberg coupling, the Dzyaloshinski-Morya (DM) interaction
and the anisotropic exchange. These correspond to the trace, antisymmetric and symmetric-traceless 
parts of the spin-spin interaction matrix, respectively. Let us focus on the bond between sites
1 and 2 (see Fig.~\ref{fig:DM_vectors}), as the spin interactions for all other bonds can be determined using the crystal symmetries.
We express the hopping Hamiltonian between these two sites as 
\begin{align}
  H_t=-c_{1\al}^\dag h_{\al\be}c_{2\be}-c_{2\al}^\dag h_{\al\be}^\dag c_{1\be}
\end{align}
where $h_{\al\be}$ is a 2 by 2 complex matrix. 
Time-reversal symmetry restricts the matrix elements as follows:
\begin{align}\label{eq:gen-hop-matrix}
  h=t\sigma^0+i\b v\cdot \b\sigma
\end{align}
where $t$ and $\b v$ are real, and $\sigma^0$ is the identity matrix. We note that in order to derive
the spin Hamiltonian, \req{eq:spin-hamil}, we want to use the same quantization axes for both sites, i.e.
we want the spin operators to be defined in the same coordinate system.

Given the hopping matrix in the form \req{eq:gen-hop-matrix}, it can be shown quite simply that the
Heisenberg, DM and anisotropic terms read
\begin{align}
  J \frac{U}{4} &=t^2-v^2/3 \\
  \b D \frac{U}{4} &= 2t\b v \\
  \Gamma^{ab}\frac{U}{4} &= 2(v^av^b -\de^{ab}v^2/3)
\end{align}

If we turn to our microscopic hopping Hamiltonian, for the $(1,2)$ bond we get
\begin{align}
  t &= a+bt_\sigma \label{eq:t}\\
  \b v &=v_y (0,1,-1) \qquad {\rm with}\; v_y=a'+b' t_\sigma \label{eq:v}
\end{align}
where we have set $t_{\rm oxy}=1$ and $t_\pi=-2t_\sigma/3$, as above.
The coefficients $a,b,a',b'$ are positive rational numbers:
\begin{alignat}{2}
a&=130/243 \approx  0.53 \quad & b &=785/2916 \approx   0.27 \\
a'&=28/243\approx  0.12 \quad & b'&=125/729 \approx  0.17
\end{alignat}
We note that the $\b v$ vector, hence $\b D$, is parallel to the opposite
bond, $(3,4)$, see Fig.~\ref{fig:DM_vectors}. This is a generic property of the pyrochlore lattice:
as a consequence of crystal symmetry, a $\b D$ vector for any given bond
must be parallel to its opposite bond (in the sense that the 4 sites form a
tetrahedron)\cite{mc}. Moreover, if we know the DM vector for
a single bond, crystal symmetries determine the DM vectors for \emph{all} other bonds 
in the lattice. Hence, there are only two possible sets of DM vectors $\{\b D_{ij}\}$, called ``direct" and ``indirect". 
They are determined by
the sign of $t v_y$ for bond $(1,2)$. (We could have picked another bond as the representative of the whole set.)
The indirect (direct) type is defined as having the $\b D$ vector
for the bond between sites 1 and 2
point along $\pm(0,1,-1)$. See Fig.~\ref{fig:DM_vectors} for the configuration of $\b D$ vectors corresponding to
the indirect DM interaction.

The nearest neighbour Heisenberg model together with a DM term on the pyrochlore lattice was studied by
classical Monte Carlo and mean field methods\cite{mc}. First, the Monte Carlo
study predicted a $\b q=0$ ordering, justifying the Ansatz used in the main text. Second, it was found that different
magnetic orders arise depending on whether
the DM interaction is of direct or indirect type.  For the direct type, the configuration was found to be unique (up
to time-reversal):
the all-in/out order mentionned above. Whereas for the indirect type, a
continous manifold of degenerate orders was found, containing both coplanar and non-coplanar 
configurations. 

For the bond $(1,2)$, we can extract from our microscopic Hamiltonian the value 
of the $\b D$ vector:
\begin{equation}
 \b D4/U=2tv_y(0,1,-1). 
 \end{equation}
 Hence, if $tv_y=(a+bt_\sigma)(a'+b't_\sigma)>0$ 
we have an indirect exchange, otherwise it is direct. It is easy to see that the $\b D$
vector changes direction when $t=0$ and $v_y=0$, which correspond to $t_\sigma\approx -1.99$
and $t_\sigma \approx-0.67$, respectively. For $t_\sigma$ between these values, the DM interaction is
of direct type, otherwise it is indirect. The behaviour of the DM interaction as a function of
the direct hopping $t_\sigma$ is shown in Fig~\ref{fig:DM_sign}. We note that first value (-0.67) is almost equal to the value
at which the $U=0$ ground state goes from an insulator to a (semi)metal, $t_\sigma=-0.65$.
The magnetic orders we find for $t_\sigma>-0.65$ belong to the continous manifold
corresponding to the indirect DM term, while it is all-in/out when $t_\sigma<-0.65$, not
too negative. Hence, the magnetic orders we get from our mean field calculation match those
obtained in the strong coupling limit. The types of magnetic orders at intermediate $U$ are
found to be related to the type of DM interaction obtained in the large $U$ spin model.

We further note that the DM interaction becomes of indirect type for $t_\sigma<-1.99$, which
is sufficiently close to the second transition in the $U=0$ ground state, from the (semi)metal
to the TI, which happens at $t_\sigma=-1.67$. For $t_\sigma<-1.67$, we get again the
magnetic orders expected for an indirect DM interaction, again consistent with the large $U$ limit.
The bigger discrepency between the point at which the $\b D$ vector changes sign and the 
value of $t_\sigma$ at which we observe a different ordering is probably due to the fact that
the anisotropic exchanges increases in importance as $t_\sigma$ is increased, while it is 
smaller than the DM interaction near the first transition in the vicinity of $t_\sigma=0$. Hence, in that regime,
we do not expect
as good of an agreement with a spin model neglecting anisotropic exchange.

\section{Discussion} 
We have constructed a minimal (but sufficiently realistic) model to describe novel quantum ground 
states that may arise in the pyrochlore iridates. While not appreciated in previous works, 
it is shown that the inclusion of both indirect and direct hopping 
process of $5d$ electrons of Ir is important in describing different magnetically ordered states in the presence
of interactions and their parent non-interacting ground states. 
A portion of our phase diagram 
is broadly consistent with a recent \emph{ab initio} calculation\cite{wan}, 
where upon increasing $U$, one encounters a metal, a topological semi-metal in the all-in/out 
magnetic configuration and finally a magnetic insulator.
Since different choices of $A$-site ions in A$_2$Ir$_2$O$_7$ lead to changes in both hopping
amplitudes, our results suggest that different magnetic and topological 
ground states such as a topological insulator, the all-in/out and related AF states and various kinds of topological semi-metals, 
may arise in a variety of pyrochlore iridates.
High pressure experiments on these compounds may reveal the intimate connection
between the magnetic order in the stronger correlation regime and TI/metal in the weak
correlation limit, as theoretically explored in this work.
For instance, recent transport measurements under high pressure\cite{fazel} on Eu$_2$Ir$_2$O$_7$ indicate 
a continuous transition from an insulating ground state to a metallic one, mimicking chemical pressure\cite{maeno}. 
This could be connected to our continuous TSM-metal transition. 
%It would be interesting to understand to what degree the observed transition
%is correlation driven versus structural. 
Also, as the existence of the TSM depends crucially on the magnetic order, 
it would be desirable to examine the effect of the magnetic fluctuations near the (semi-)metal-TSM transition on thermodynamic and transport properties.

\section*{Acknowledgements}
We are grateful to S. Bhattacharjee, G. Chen, A. Go, S. R. Julian, Y. J. Kim, D. E. MacLaughlin, S. Nakatsuji, D. Podolsky, J. Rau,  T. Senthil, F. Tafti, H. Takagi, A. Vishvanath and B. J. Yang for useful discussions. WWK acknowledges the hospitality of the Korea Institute for Advanced Study and MIT where parts of the research were done. This work was supported by NSERC, FQRNT, the CRC program, and CIFAR.
\bibliography{pyro_r1}{}

%\begin{thebibliography}{99}
%\bibitem{montecarlo} M. Elhajal, B. Canals, R. Sunyer, and C. Lacroix, Phys. Rev. B 71,
%094420 (2005).
%\end{thebibliography}
\end{document}